\documentstyle[preprint,pra,aps,epsf]{revtex}
\begin{document}
\draft
\title{\bf
Limits on the monopole magnetic field from measurements of the
electric dipole
moments of atoms, molecules and the neutron}
\author{V. V. Flambaum$^{1,2,}$
\footnote{e-mail: flambaum@newt.phys.unsw.edu.au}
and D. W. Murray$^1$}
\address{$^1$School of Physics, University of New South Wales,
Sydney, 2052, Australia\\
$^2$  ITAMP, Harvard University and the Smithsonian Astrophysical
Observatory,
60 Garden Street, Cambridge, Massachusetts 02138}
%\date{\today}
\maketitle
\begin{abstract}
A radial magnetic field can induce a time invariance violating
electric
dipole moment (EDM) in quantum systems. The EDMs of the Tl, Cs,
Xe and Hg
atoms and the neutron that are produced by such a
field are estimated.
The contributions of such a field to the constants, $\chi$ of
the T,P-odd interactions $\chi_e {\bf N} \cdot {\bf s}/s$
and $\chi_N {\bf N} \cdot {\bf I}/I$ are also estimated
for the TlF, HgF and YbF molecules (where ${\bf s}$ (${\bf I}$)
is the electron (nuclear) spin and ${\bf N}$ is the
molecular axis).
The best limit on the contact
monopole  field can be
obtained from the measured value of the
Tl EDM. The possibility of such a field being produced from
polarization of the
vacuum of electrically charged magnetic monopoles (dyons) 
by a Coulomb field is discussed, as well as
the limit on these dyons. An alternative mechanism involves
 chromomagnetic and chromoelectric fields in QCD.\\
% Submitted to PHYS. REV. A
\end{abstract}
\vspace{5mm}
\pacs{PACS number(s): 14.80.Hv, 11.30.Er, 32.10.Dk}
\vspace{10mm}

\section{Introduction}

Dirac \cite{Dirac1931} considered magnetic monopoles and
derived a
quantization rule for magnetic
charge $M$: $eM= \frac{\hbar c}{2}k$,
where $k$ is an integer (below we put $\hbar=c=1$). Zwanziger and
Schwinger \cite{Zwanziger1968,Schwinger668} generalized
this condition for dyons, which carry both electric ($q$) and
magnetic charges: $q_1 M_2 - q_2 M_1 = \frac{k}{2}$. This
formula may be derived
heuristically by the quantization of the angular
momentum (half-integer
or integer) of the electromagnetic field in a system
consisting of two dyons
\cite{Saha1936,Fierz1944,Wilson1949,Sokolov1976,Sokolov1977}.
E. Purcell and N. Ramsey in ref. \cite{PurcellRamsey} and
N. Ramsey \cite{Ramsey} discussed the possibility of
there being elementary particle
and nuclear EDMs due to the existence of magnetic monopoles.
Later, E. Witten \cite{Witten1979} showed
that 't Hooft-Polyakov magnetic
monopoles carry small electric charges due
to CP violating interactions.
In the case of the $\theta$-term (the
interaction $\theta \frac{e^2}{32 \pi^2}
F_{\mu \nu} \tilde{F}_{\mu \nu}$) the electric charge of the
monopole is $q=-\frac{e \theta}{2 \pi}$.
V. Sokolov developed nonrelativistic
Lagrangian and Hamiltonian formalisms
for the interaction of electric and magnetic charges which do
not involve ``strings'' \cite{Sokolov1976,Sokolov1977}.

It is possible that monopoles may appear in particle-antiparticle
pairs in which
the point-like positive and negative magnetic charges could
be very hard or impossible to
separate. Recall that a ``supercritical'' electric
charge $Ze$ with $Z>1/ \alpha$ must be screened down to
the value $Z=1/ \alpha$  due to spontaneous $e^+ e^-$
pair production. The strength of
the interaction in this case ($Ze^2 = Z\alpha \sim 1$) is
still smaller
than $M^2 \sim 1/ \alpha$ (using the quantization rule for
magnetic charge).
Also, Nambu showed that in the standard
electroweak model the classical solution is a
monopole-antimonopole
pair connected by a $Z^0$-field string \cite{Nambu1977}. This
could explain the absence of free monopoles. However even
in this case one can still search for the effects of
virtual monopole-antimonopole pairs
and this will be discussed in the present paper.

In ref. \cite{Flambaum1994} it was pointed out that an
electric dipole moment (EDM) of an atom (or any quantum system)
can be induced by the interaction of the electrons
with a radial magnetic
field (${\bf B} \propto {\bf r}$). The existence of such
a field would contradict both time reflection invariance (T)
and Gauss's Law ($\oint {\bf B} \cdot d{\bf a} = 0$ ). The
limit on such a field can be of interest by itself as a
very precise test of electrodynamics at small distances since
the accuracy of EDM measurements is very high now.
It was also pointed out in \cite{Flambaum1994} that a
radial magnetic field can be produced due to a
monopole-antimonopole pair contribution to the magnetic
moment (if there is no magnetic string) and a time invariance
violating interaction which polarizes nucleon spins
along the radial direction.

The mechanisms of radial magnetic field and EDM creation
can be much simpler if the magnetic monopole has an
electric charge or is subject to the strong interaction.
The simplest example would be for a  magnetic
charge to be captured by an atomic nucleus and hence produce a
``Coulomb'' magnetic field ${\bf B} = \frac{M {\bf r}}{r^3}$.
It is easy to find the
exact solution of this problem (an electron in the field
of the dyon) in the nonrelativistic case. All stationary states
in this problem possess an EDM.
 Another example is a dyon-antidyon system with nonzero
orbital angular momentum.
Similarly to the way in which orbiting electric charges produce a
magnetic moment, orbiting magnetic charges produce
an EDM \cite{PurcellRamsey}.

The most interesting possibility would be for the radial
magnetic field and EDM to be induced by the virtual
production of dyon-antidyon pairs.  The mechanism could be the
following.  The electric field of
an atomic nucleus polarizes the vacuum of dyon pairs and creates
corrections to the radial electric field, $\delta {\bf E}$. The
ratio of the magnetic to the electric field produced by
these dyons is $\frac{{\bf B}}{{\bf E}} = \frac{M}{q}$.
Thus dyon vacuum polarization produces the radial magnetic
field ${\bf B} = \frac{M}{q} \delta {\bf E}$.
The interaction of atomic electrons with this
magnetic field produces an
atomic EDM. A similar mechanism could produce 3--5 orders
of magnitude bigger P,T-odd effects in diatomic polar
molecules. We also consider the
contribution of this mechanism to the neutron EDM in the
constituent quark model. Note that the electric field can
be replaced by the strong field if the monopole interacts
strongly. Moreover, it seems that ``chromodyons'',
which could exist in a generalization of QCD, could produce
similar effects to that of ``electromagnetic'' dyons.
Thus the problem
considered in the present work could be related to the recent
ideas
about the role of monopole condensate in quark confinement
(see e.g. the recent review in \cite{PhysicsToday} and
references therein).

In this paper we calculate the possible effects of time invariance
violation in atoms, molecules and the neutron produced by a
radial magnetic field. We also estimate the field
produced by the polarization of the dyon vacuum according
to the mechanism discussed above.
An accurate calculation of the monopole effects requires
the solution of numerous complicated problems such as the
large value of the magnetic charge, ``strings'', the finite
size of the classical monopole solution, etc. We stress that
in the present work we are mainly trying to avoid these
problems rather than to solve them. In fact we explore an
approach using simple heuristic arguments and perturbation
theory which allows us to estimate and compare the values of
T,P-odd effects in different quantum systems. We must
add that our attempt to use the results of
two-vector potential theory
for the dyon electrodynamics \cite{pan}
(see also \cite{Zwanziger1971,Brandt})
has lead to the conclusion that the magnetic field due
to dyon vacuum polarization seems to vanish within this
theory. (This contradicts the simple and natural
picture discussed above! Note however that the theory
\cite{pan} includes dyons only and the introduction of
the usual charges into this theory leads to
serious complications.)
An alternative approach is to use the theories
with 't Hooft-Polyakov magnetic monopoles. We have
not done any calculations within these theories.
Our more simple ``heuristic'' calculation
of the radial magnetic field due to 
dyon vacuum polarization (see below) is not based on
a complete consistent theory and strictly speaking does
not prove the existence of the effect.
However, the calculations of the effects of this field in
sections \ref{sc}--\ref{se} apply to any contact radial
magnetic field; they are not restricted to fields produced
by dyon vacuum polarization. Thus, the results in these
sections can be used in general by using the equations
in terms of $B_0$ ($B_0$ is defined below).

\section{The radial magnetic field due to dyon vacuum polarization}

Let us start from an estimate of the radial magnetic field
which could be produced due to polarization of the dyon
vacuum by an electric charge. The correction to the
electrostatic potential $\frac{e_1}{r}$ due to the
vacuum polarization of spin $\frac{1}{2}$ particles can be
found in any
textbook on quantum electrodynamics (see e.g. \cite{Ber.L.P.}):
\begin{equation}
\label{e1}
\Phi(r) = \frac{e_1}{r}
\frac{2 q^2}{3 \pi} \int_1^{\infty} e^{-2mr \zeta}
\left(1+\frac{1}{2\zeta^2}\right)
\frac{\sqrt{\zeta^2-1}}{\zeta^2} \, d\zeta \; ,
\end{equation}
where $q$ and $m$ are the electric charge and
mass of the dyons. (In fact, $2m$ in eq. (\ref{e1}) is the
threshold of production of a dyon pair. In the case of a bound
pair (connected e.g. by a $Z^0$-string) we could substitute the
mass of the two-dyon system instead.) The magnetic field due
to the dyons can be expressed in terms of the
corresponding electric
field: ${\bf B} = {\bf E}\cdot
\frac{M}{q}= -\mbox{\boldmath $\nabla$} \Phi\cdot\frac{M}{q}$,
where $M$ is the magnetic charge. If the dyons are heavy then
the potential can be written as $\Phi({\bf r}) =
{\rm constant}\cdot\delta({\bf r})$, where
the constant can be found by the integration of
eq.~(\ref{e1}) over
${\bf r}$. Thus polarization of the dyon vacuum produces the
following
radial magnetic field around the point-like charge $e$:
\begin{equation}
\label{e2ex1}
{\bf B} \equiv B_0 \cdot \mbox{\boldmath $\nabla$}
\delta({\bf r}) \; ,
\end{equation}
with
\begin{equation}
\label{e2ex2}
B_0 = -\frac{4}{15} \frac{eqM}{m^2} \; .
\end{equation}  
Note that T-invariance in this case can be restored if there
is one more dyon
with the same mass but the opposite sign of the product $qM$.

We can obtain the radial magnetic field of the nucleus by
replacing $\delta ({\bf r})$ by the proton density
distribution, $\rho_p(r)
\approx Z \rho_0 \cdot \theta(R-r)$:
\begin{equation}
\label{e3}
{\bf B} = B_0 \cdot \mbox{\boldmath $\nabla$}
\rho_p(r) \approx -Z\rho_0
B_0 \delta(r-R) {\bf n} \; ,
\end{equation}
where ${\bf n} = {\bf r} / r$ is a radial unit vector
and $\rho_0 = (\frac{4}{3} \pi R^3)^{-1}$. Here we
took into account the fact that the nuclear
density varies in a small interval around the nuclear radius $R$.

\section{The interaction between an electron and the
radial magnetic field}
\label{sc}

Consider now the interaction between an atomic electron and the
contact radial
magnetic field. A radial magnetic field cannot be described
 by a nonsingular vector potential ${\bf A}$. Therefore, we will
avoid using the vector potential. Let us start from the
nonrelativistic problem. The interaction between the spin magnetic
moment and the magnetic field does not contain the vector potential:

\begin{equation}
\label{nonrel}
V = - \mu \bbox{\sigma} \cdot {\bf B} = Z \mu B_0 \rho_0
   \bbox{\sigma} \cdot {\bf n} \delta(r-R) \; ,
\end{equation}
where $\mbox{\boldmath $\sigma$}$ are Pauli matrices. For
the electron
$\mu = -\frac{e}{2m_e}$, with $e>0$.
Note that $B_0$ in the above equation is only given by
eq. (\ref{e2ex2}) for the dyon vacuum polarization mechanism,
and can otherwise be considered as a more general parameter.

The orbital contribution
to the interaction seems to vanish due to a cancelation
between the contributions of the radial magnetic field
of the nucleus and that of the electron. Let us first
calculate the force acting 
on the electron from the radial magnetic field of the nucleus,
in the latter's rest frame:
\begin{equation}
\label{force}
{\bf F}_1 = -e ({\bf v} \times {\bf B}) /c \; ,
\end{equation}
where $ {\bf v}$ is the electron's velocity.
To calculate the second force acting on the electron, we first
calculate the force acting on the nucleus from the magnetic field
of the electron ${\bf B}_e$. Since the electron is
in motion, this magnetic field will be transformed into an
electric field in the rest frame of the nucleus:
\begin{equation}
\label{electric}
{\bf E} = - ( {\bf v} \times {\bf B_e}) /c \;
\end{equation}
(to terms of order $v/c$).

Let us assume
for now that the radial magnetic field
around a particle is proportional to its electric
charge.
This is obviously true for the magnetic field due to
the dyon vacuum polarization mechanism in 
eqs. (\ref{e2ex1}) and (\ref{e2ex2}), however this discussion
is intended to be more general. 
(An argument
for this proportionality that is based on angular
momentum quantization will be given below.)
 Under this assumption the force on the nucleus
from the electric field (\ref{electric}) is exactly equal
to ${\bf F}_1$ in eq. (\ref{force}). According
to Newton's third law
the corresponding force acting on the electron from the nucleus is
${\bf F}_2=-{\bf F}_1$, i.e. the  net force 
acting on the electron is
$ {\bf F}= {\bf F}_1+{\bf F}_2 = {\bf 0}$. One can say that a
cancelation occurs between the force  
${\bf F}_1$ acting on the electron from
the vacuum monopole distribution near the nucleus
and the force  ${\bf F}_2$ acting
on the vacuum monopole distribution near the 
electron due to the electric field of the nucleus.

There is a second argument in favor of the
cancelation of the orbital
contribution. As is known
\cite{Saha1936,Fierz1944,Wilson1949,Sokolov1976,Sokolov1977}, the
radial magnetic field of a magnetic
charge and the radial electric field of an electric charge
together produce a nonzero
angular momentum  around the axis connecting the 
charges ($\sim \bf{E}\times\bf{B}$). In Quantum
Mechanics angular momentum is quantized. This requirement 
implies the
quantization of the product of the magnetic 
and the electric charges.
For two dyons this 
condition is  $K=q_1 M_2 - q_2 M_1 = \frac{k}{2}$,
where $K=\bf{J}\cdot
\bf{n}$ is the projection of the
angular momentum $\bf{J}$ onto the axis connecting
these particles and $\bf{n}$ is a unit vector along this axis.
Recall that the system's orbital 
angular momentum is orthogonal to the connecting axis and
does not contribute to $K$.
We see that for $k=0$ the electromagnetic field angular 
momentum is zero due to the cancelation between 
the contributions of these two dyons.

A similar situation arises for the short-range induced
magnetic field. If both the radial magnetic fields of
the electron and the point-like nucleus
are proportional to their charges then 
the angular momentum of the field will be zero due to 
a cancelation between the two contributions.
(Actually, the nucleus is not point-like, therefore, strictly
speaking, we should consider an electron-quark system here.)
We stress that in the absence of such a
cancelation there would be a problem: the size of the 
magnetic field region
is very small ($\sim m^{-1}$), and so we cannot satisfy 
the condition of
angular momentum quantization for nonzero $K$ (this provides an
argument for the proportionality of the radial magnetic
fields to the electric charges).

Now there is a relation
between the orbital contribution to the
EDM of the two-particle system and $K$. 
The angular wave function for the system
is the Wigner $D$-function, $D^J_{MK}(\bf{n})$, where $M$ is the
 projection of  ${\bf J}$ onto the z-axis  
(compare e.g. with ref. \cite{Sokolov1976}, where the 
charge-monopole solution was found).
The orbital contribution to the
EDM in such an angular state is
proportional to the angular integral
$\int |D^J_{MK}({\bf n})|^2 {\bf n}_z \, d\Omega=K M/[J(J+1)]$. 
This relation is especially simple for $K=0$ since
the $D$-function in this case
coincides with the usual $Y_{LM}$ angular function which gives an
(orbital) EDM of zero.
Thus, we can consider the zero value of the
electromagnetic field angular momentum in
the electron-nucleus case to be
an argument for the absence of an orbital contribution 
to the electric dipole moment.

We must stress that there is no such cancelation between
the contributions of the magnetic moments of the nucleus and
the electron (see eq. (\ref{nonrel})) since
these magnetic moments are very different. For example,
the nuclear magnetic moment can be zero.

In the relativistic case the interaction of
a magnetic moment
$\mu$ with a magnetic field can be expressed in terms of
the magnetic field only using the following well known
identity (see e.g.
\cite{Ber.L.P.}):
\begin{equation}
\label{e4}
j^{\mu} = \overline{\psi}_2 \gamma^{\mu} \psi_1
 = \frac{1}{2m} \overline{\psi}_2 ({p_{1}}^{\mu} + {p_{2}}^{\mu})
\psi_1 - \frac{1}{2m} \overline{\psi}_2 \sigma^{\mu \nu} k_{\nu}
\psi_1 \; ,
\end{equation}
where the first term in the r.h.s. is an orbital contribution
to the electromagnetic current and the second term is
a spin one and $k_{\nu} =
p_{2\nu}-p_{1\nu}$. Taking into account
the fact that $i({\bf k} \times {\bf A})
= {\bf B}$ we obtain (from
$\langle \psi_2 |V|\psi_1 \rangle=
-e \langle \psi_2 | j^{\mu} A_{\mu} | \psi_1 \rangle$) 
a relativistic expression for the 
interaction of a
magnetic moment $\mu$ with the radial magnetic field (\ref{e3}):
\begin{equation}
\label{e5}
V = - \mu \beta {\bf \Sigma} \cdot {\bf B} = Z \mu B_0 \rho_0
    \left( \begin{array}{cc}
            \mbox{\boldmath $\sigma$}\cdot{\bf n} & 0 \\
            0 & - \mbox{\boldmath $\sigma$}\cdot{\bf n}
           \end{array}
    \right) \delta(r-R) \; .
% \; ,
\end{equation}
Expression (\ref{e5}) can also be
obtained from the quadratic form of the Dirac equation.

\section{The atomic electric dipole moment and T,P-violation 
in molecules}
\label{sd}

In this section we calculate the contribution of dyons to the
atomic EDM, as well as their effect on molecules.
The interaction (\ref{e5}) mixes atomic electron states of
opposite
parity, mostly ${\rm s}_{\frac{1}{2}}$ and ${\rm p}_{\frac{1}{2}}$
orbitals which are large at the nuclear surface.
\begin{eqnarray}
\psi_{{\rm s}_{\frac{1}{2}}} (R) = \left( \begin{array}{c} f_s \\
-i(\mbox{\boldmath $\sigma$} \cdot {\bf n}) g_s \end{array}
\right)
\frac{\chi}{\sqrt{4\pi}} &
\;\;\;\; , \;\;\;\;&
\psi_{{\rm p}_{\frac{1}{2}}} (R) = \left( \begin{array}{c}
-(\mbox{\boldmath $\sigma$} \cdot {\bf n})f_p \\ i g_p \end{array}
\right) \frac{\chi}{\sqrt{4\pi}}
\nonumber \\
f_s f_p & \approx & N_0 \cdot \left[Z^2 \alpha^2 + \frac{4}{3}
\frac{Z R}{a}\right] \nonumber \\
g_s g_p & \approx & -N_0 Z^2 \alpha^2 \nonumber \\
N_o & = & \frac{Z R_r}{a^3 (\nu_s \nu_p)^{\frac{3}{2}}}
\nonumber \\
\label{e5b}
R_r & = & \left[ \frac{2}{\Gamma(2\gamma+1)} \left( \frac{a}{2ZR}
\right)^{1-\gamma} \right]^2
\end{eqnarray}
We use expressions for the electron wave functions at the nuclear
surface from \cite{Khrip}. In the notation used in this book
$E=-\frac{e^2}{2a\nu^2}$ is the electron
energy, $a=({m_e e^2})^{-1}$ is
the Bohr radius, $\Gamma$ is the gamma function,
$\gamma = \sqrt{1-Z^2
\alpha^2}$ and $\chi$ is the electron spinor. Using
$(\mbox{\boldmath
$\sigma$} \cdot {\bf n})^2 = 1$ we obtain the matrix element
of the interaction of the electron
magnetic moment with the radial magnetic
field ${\bf B}$ (\ref{e5}):
\begin{equation}
\label{e6}
\langle {\rm s}_{\frac{1}{2}} | V | {\rm p}_{\frac{1}{2}}
\rangle =
\frac{3e}{4\pi}
\frac{Z(Z\alpha)R_r B_0}{a^2(\nu_s \nu_p)^{\frac{3}{2}}}
\left[ \frac{Z^2 \alpha^2}{R} 
+ \frac{2}{3} \frac{Z}{a} \right] \; .
\end{equation}
Note that the first term in the brackets is much larger in
heavy atoms
and the second term is necessary for the correct
nonrelativistic limit
as $Z\alpha \rightarrow 0$. (It is interesting that the
``relativistic enhancement factor'' in this case can
exceed several hundred.)

The electric dipole moment of an atom with one external
electron (e.g. Tl, Cs, Fr) generated by the
interaction (\ref{e6}) can be calculated
using perturbation theory in $V$:
\begin{equation}
\label{e7}
d_A = \langle \tilde{\psi} | -e {\bf r} | \tilde{\psi} \rangle =
\frac{2}{3} e \sum_n \frac{r_{0n} V_{n0}}{E_0-E_n} \; ,
\end{equation}
where $\tilde{\psi}$ is the perturbed wave function, $r_{0n}$ is
the radial integral
($\langle {\rm s}_{\frac{1}{2}} | r_z | {\rm p}_{\frac{1}{2}}
\rangle = -\frac{1}{3} r_{sp}$) and $V_{n0}$ is the matrix
element of $V$ between the $|{\rm s}_{\frac{1}{2}} \rangle$ and
$|{\rm p}_{\frac{1}{2}} \rangle$  orbitals
($|0\rangle = |6 {\rm s}_{\frac{1}{2}} \rangle$
and $|n\rangle = |n {\rm p}_{\frac{1}{2}} \rangle$
 in Cs, $|0\rangle = |6 {\rm p}_{\frac{1}{2}} \rangle$,
$|n\rangle = |n {\rm s}_{\frac{1}{2}} \rangle$ in Tl). There is a
simpler way to obtain numerical results for the atomic EDM: to use
existing calculations of the
electron EDM enhancement factor in atoms
and molecules \cite{Khrip,Flambaum1976,KozLab95,KozEz94} or
calculations of the atomic EDM produced by the T,P-odd
electron-nucleon interaction 
$V_{ps} = i \frac{G}{\sqrt{2}} k_{1N} \overline{\psi}_e
\gamma_5 \psi_e \overline{\psi}_N \psi_N$, where $k_{1N}$ is a
dimensionless interaction constant ($k_{1p}$ or $k_{1n}$)
\cite{Khrip,KozLab95,KozEz94,HLS76}. Comparison of the matrix
element of the effective interaction between the electron EDM
$d_e$ and the atomic electric field ${\bf E}$ ($-d_e (\beta-1)
{\bf \Sigma} \cdot {\bf E}$ \cite{Khrip,Flambaum1976}) with
expression (\ref{e6}) as well as a similar comparison with the
matrix element of the T,P-odd interaction $V_{ps}$ give
the following substitutions in the expressions
for the atomic and molecular EDM
(and linear energy shifts in external electric fields):
\begin{displaymath}
\frac{d_e}{e\cdot {\rm cm}} \rightarrow \left( \frac{{\rm TeV}}
{\widetilde{m}} \right)^2 \times -3.92 \cdot 10^{-25} \frac{Z R_r
\gamma (4\gamma^2-1)}{A^{\frac{1}{3}}}
\approx \left( \frac{{\rm TeV}}{\widetilde{m}} \right)^2 \cdot
\left\{ \begin{array}{ll}
                         -6 \cdot 10^{-23} & \mbox{for Tl, Hg, 
TlF,}\ldots \\
                -2.6 \cdot 10^{-23} & \mbox{for Cs, Xe,}\ldots
        \end{array}
\right.
\end{displaymath}
or
\begin{equation}
\label{e8}
k_{1p} \rightarrow -0.87 \cdot 10^{-3}
\frac{Z}{\gamma A^{\frac{1}{3}}}
\left( \frac{{\rm TeV}}{\widetilde{m}} \right)^2 \approx \left(
\frac{{\rm TeV}}{\widetilde{m}} \right)^2 \cdot
\left\{ \begin{array}{ll}
                         -1.5 \cdot 10^{-2} & \mbox{for Tl, Hg,
TlF, HgF,}\ldots \\
                 -1.0 \cdot 10^{-2} & \mbox{for Cs, Xe,}\ldots
        \end{array}
\right.
\end{equation}
\begin{displaymath}
\widetilde{m}^2 = \frac{m^2}{q M} = -\frac{4e}{15 B_0} \; .
\end{displaymath}
To avoid confusion note that the limits are usually presented for
$C^{sp} \equiv k_1 = 0.4 k_{1p} + 0.6 k_{1n}$.

Using equation (\ref{e8}) and the results of numerical atomic and
molecular calculations of T,P-odd effects induced by the electron
EDM or the interaction $V_{ps}$ (see e.g.
\cite
{Khrip,Flambaum1976,KozLab95,KozEz94,HLS76,FlKhJETP1985,VF,CSH91})
one can easily calculate the contribution of dyons to the atomic
EDM and the constants, $\chi$ of the T,P-odd interactions
$\chi_e {\bf N} \cdot {\bf s}/s$ and
$\chi_N {\bf N} \cdot {\bf I}/I$ for molecules
(here ${\bf s}$ is the electron and ${\bf I}$ the nuclear spin
and ${\bf N}$ is the molecular axis). See tables \ref{t1}
and \ref{t2}.
Note that for the atoms and molecules with closed electron shells
(Hg, Xe, TlF) the effect is proportional to the hyperfine
interaction. For these we present rough estimates based on 
the expression for $k_{1p}$ in eq.~(\ref{e8}). However a
more accurate calculation is possible using the approach of
ref.~\cite{FlKhJETP1985}.

One can also use eq.~(\ref{e8})
to calculate the limit on $\widetilde{m}^2$ from
the known limits on
the electron EDM $d_e$ and $k_{1p}$. At present the
best limit follows
from the measurement of the EDM of the Tl atom
\cite{Commins1994}: $d_A({\rm Tl}) = [-1.05 \pm 0.70 \pm 0.59]
\cdot 10^{-24} e\cdot\mbox{ cm}$ or $d_e = [1.8 \pm 1.2 \pm 1.0]
\cdot 10^{-27} e\cdot\mbox{cm}$. Using eq.~(\ref{e8}) or
table \ref{t1} we obtain the
following limits for the dyon mass $m$
and the radial magnetic field produced by a particle
with charge $e$
(${\bf B} = B_0 \mbox{\boldmath $\nabla$} \delta ({\bf r})$):
\begin{equation}
\label{e9}
\frac{1}{\widetilde{m}^2}
\equiv \frac{qM}{m^2} = -\frac{15B_0}{4e} =
\frac{1}{[100\mbox{ TeV}]^2} \cdot [0.35 \pm 0.23 \pm 0.20] \; ,
\end{equation}
i.e. $|\widetilde{m}|=|\frac{m}{\sqrt{qM}}| > 100\mbox{ TeV}$.
According to Dirac \cite{Dirac1931} $e M = \frac{1}{2}$ and if
the dyon charge $q \sim e$ the dyons would be very heavy.
The situation is different if the product $qM$ is proportional
to the strength of the T-violating interaction. Recall that
according to \cite{Witten1979} $|qM| \approx \frac{\theta}{4\pi}$
and the present limit is $|\theta| < 4 \cdot 10^{-10}$
\cite{Smith1990,Hg1993}. In this case $qM < 30 \cdot 10^{-12}$
and so $m > 100\mbox{ MeV}\cdot \sqrt{qM \cdot 10^{12}}$ is
not necessarily large. Note also that
according to \cite{Nambu1977} the mass of a
monopole-antimonopole pair (connected by a $Z^0$-string) in the
standard model is in the TeV range.

\section{The neutron electric dipole moment}
\label{se}

Another way to search for T-violation is by neutron EDM
measurement. We can easily estimate the neutron EDM using
eqs. (\ref{e2ex1})--(\ref{e7}) and the
constituent quark model in a 3D-oscillator
potential. The magnetic moment of the neutron can be
reproduced if we assume that the quarks have
magnetic moments $\mu_q = \frac{e_q}{2m_q}$,
masses $m_q = \frac{m_n}{3}$ and are in 1s states with
the  total spin
of the d-quarks equal to 1 (SU(6) model).
The matrix element $\langle
{\rm s}_{\frac{1}{2}} | V | {\rm p}_{\frac{1}{2}} \rangle$ in
eq.~(\ref{e7}) can be easily estimated using the values of the
oscillator wave functions at $r=0$, $\psi_s (0)$ and $\psi_p (0)$.
$V$ in this case is the interaction between the magnetic moment of
one quark and the radial magnetic field produced by the
two other quarks (with total charge $e_2 + e_3 = -e_1$). The
sum over the
excited states in eq.~(\ref{e7}) is saturated by the nearest state
$1{\rm p}_{\frac{1}{2}}$ in the oscillator model and the
result can be expressed in terms of one
parameter, $a_n = \sqrt{\frac{\hbar}{2 m_q \omega}}$,
which is in fact the size of the
nucleon ($a_n \sim \frac{1}{m_{\pi}}$,
where $m_{\pi}$ is the $\pi$-meson mass).
After summation over the 3 quarks the EDM of the neutron is:
\begin{equation}
\label{e10}
d_n \sim \sum_{q=1}^{3} \frac{2 e_q^3 \cdot (B_0 / e)}{3
(2\pi)^{\frac{3}{2}} a_n} \cdot \langle \sigma_z^{(q)} \rangle
= 1.7 \cdot 10^{-3} \frac{\alpha e}{\widetilde{m}^2 a_n}
\sim 3 \cdot 10^{-26} e\cdot \mbox{cm} \left( \frac{\mbox{TeV}}
{\widetilde{m}} \right) ^2 \; .
\end{equation}
Of course this is only an order of magnitude estimate which is
however good enough to say that the neutron
EDM is $10^5$ times smaller than that of a heavy atom with
nonzero electron angular momentum (see table \ref{t1}). The
electron EDM will be even smaller since it
should be proportional to $\frac{m_e}{\widetilde{m}^2}$
and $m_e \sim 10^{-3} m_n$.

A neutron EDM can also be made up of ``intrinsic'' quark EDMs.
This quark
EDM (as well as the electron's EDM) could appear due to radiative
corrections with dyon loops. See Fig. 1.
Fig. 1a corresponds to the vacuum transformation of
a homogeneous electric field to a magnetic
field ($F_{\mu \nu} \tilde{F}_{\mu \nu}$ term) and
should be omitted. Diagram 1b is forbidden
by the Furry theorem (which however must be checked
for dyons). Other diagrams such as that in Fig. 1c
are of a higher order in the electromagnetic interaction.
They are not
necessarily small (since $qM$ could be $\sim 1$) but they
hardly can
exceed the lower order contribution of Fig. 1d calculated in
eq.~(\ref{e10}).

We stress once more that a similar mechanism can 
possibly produce an EDM of
nucleons and nuclei if we consider ``chromomagnetic''
monopoles instead of the ``usual'' monopoles. All that
is required is to replace electric ($q$)
and magnetic ($M$) charges in the expression
for $\widetilde{m}^2 = \frac{m^2}{qM}$ by
``chromoelectric'' (color) and ``chromomagnetic''
charges. An atomic EDM in this case can
be generated by the interaction
between electrons and nuclear T,P-odd moments.

\begin{acknowledgements}

One of us (VVF) is deeply thankful to M. Yu. Kuchiev for
stimulating discussions. He is also
grateful to I.B. Khriplovich, B. Marciano, V. Soni, E. Shuryak,
O. Sushkov and participants of the program
``Physics Beyond the Standard
Model'' at the Institute of Nuclear Theory (University
of Washington) for valuable discussions and references.
This work is supported by the Australian Research Council and
by the National Science Foundation through a grant for the
Institute for Theoretical Atomic and Molecular Physics
at Harvard University and the Smithsonian Astrophysical
Observatory. 
\end{acknowledgements}

\begin{figure}
\leavevmode
\begin{center}
\begin{tabular}{c}
\epsfxsize=243pt
\epsffile{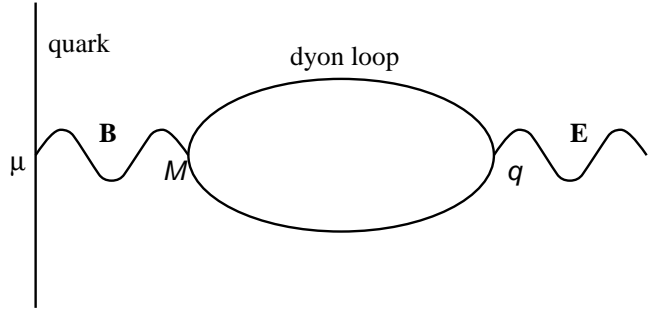}
\\
(a)
\\
\epsfxsize=243pt
\epsffile{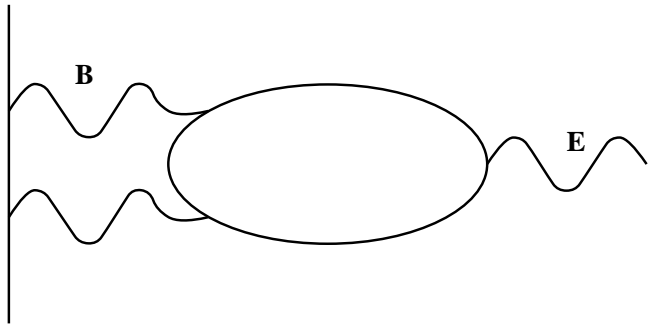}
\\
(b)
\\
\epsfxsize=243pt
\epsffile{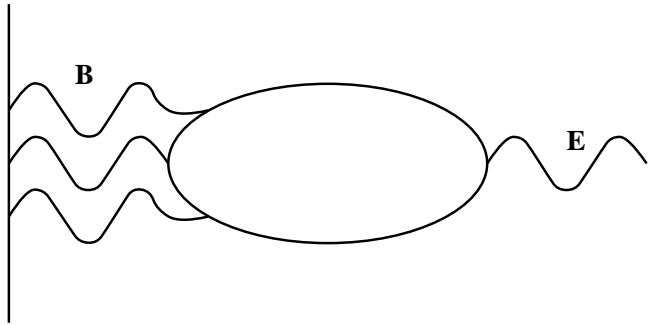}
\\
(c)
\\
\epsfxsize=243pt
\epsffile{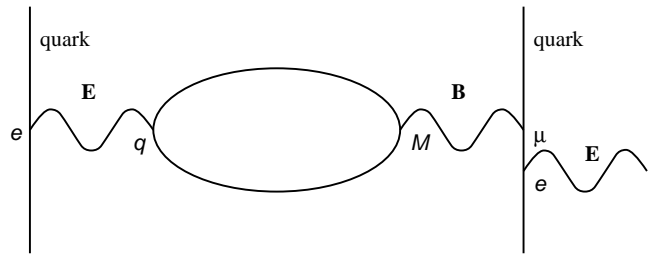}
\\
(d)
\end{tabular}
\end{center}
\caption{Diagrams showing the production of a quark
EDM due to radiative corrections with dyon loops (a--c)
and the diagram showing the mechanism calculated
in eq. \protect\ref{e10} (d).}
\end{figure}
\begin{table}
\begin{center}
\begin{tabular}{l r}
atom & $\frac{d_A}{e \cdot {\rm cm}}
\cdot \left(\frac{\widetilde{m}}{\rm TeV} \right)^2$ \\ \hline
Tl & $3\cdot10^{-20}$ \\
Cs & $-3\cdot10^{-21}$ \\
Hg & $\sim 3\cdot10^{-24}$ \\
Xe & $\sim 1.5\cdot10^{-25}$ \\
Xe ($^3 P_2$) & $-3\cdot10^{-21}$
\end{tabular}
\end{center}
\caption{The contribution of a radial magnetic field (due
to virtual dyons) to various atomic EDMs.}
\label{t1}
\end{table}
\begin{table}
\begin{center}
\begin{tabular}{l r}
molecule & $\chi \cdot \left(\frac{\widetilde{m}}{\rm TeV}
\right)^2
$\mbox{ (Hz)} \\ \hline
TlF ($\chi_N$) & $\sim -0.1$ \\
HgF ($\chi_e$) & 1500 \\
YbF ($\chi_e$) & 300
\end{tabular}
\end{center}
\caption{The contribution of a radial magnetic field (due
to virtual dyons) to the constants $\chi$ for molecules.}
\label{t2}
\end{table}  

\begin{thebibliography}{99}
\bibitem{Dirac1931}
P.A.M. Dirac, Proc. Roy. Soc. (London) {\bf A133}, 60 (1931).
\bibitem{Zwanziger1968}
D. Zwanziger, Phys. Rev. {\bf 176}, 1480; 1489 (1968).
\bibitem{Schwinger668}
J. Schwinger, Phys. Rev. {\bf 144}, 1087 (1966);
{\bf 173}, 1536 (1968).
\bibitem{Saha1936}
M.N. Saha, Indian J. Phys. {\bf 10}, 145 (1936).
\bibitem{Fierz1944}
M. Fierz, Helv. Phys. Acta. {\bf 17}, 27 (1944).
\bibitem{Wilson1949}
H.A. Wilson, Phys. Rev. {\bf 75}, 309 (1949).
\bibitem{Sokolov1976}
V.V. Sokolov, Yad. Fiz. {\bf 23},
628 (1976) [Sov. J. Nucl. Phys. {\bf 23}, 330 (1976)].
\bibitem{Sokolov1977}
V.V. Sokolov, Yad. Fiz. {\bf 26}, 427
(1977) [Sov. J. Nucl. Phys. {\bf 26}, 224 (1977)].
\bibitem{PurcellRamsey} 
E.M. Purcell and N.F. Ramsey,
Phys. Rev. {\bf 78}, 807 (1950).
\bibitem{Ramsey}
N.F. Ramsey, Phys. Rev. {\bf 109}, 225 (1958).
\bibitem{Witten1979}
E. Witten, Phys. Lett. B {\bf 86}, 283 (1979).
\bibitem{Nambu1977}
Y. Nambu, Nucl. Phys. B {\bf 130}, 505 (1977).
\bibitem{Flambaum1994}
V.V. Flambaum, Phys. Lett. B {\bf 320}, 211 (1994).
\bibitem{PhysicsToday}
G.P. Collins, Phys. Today {\bf 48} (3), 17 (1995).
\bibitem{pan}
C. Panagiotakopoulos, Nucl. Phys. B {\bf 212}, 118 (1983). 
\bibitem{Zwanziger1971}
D. Zwanziger, Phys. Rev. D {\bf 3}, 880 (1971).
\bibitem{Brandt}
R.A. Brandt, F. Neri and D. Zwanziger, Phys.
Rev. D {\bf 19}, 1153 (1979).
\bibitem{Ber.L.P.}
V.B. Berestetski\v{\i}, E.M. Lifshitz and L.P. Pitaevski\v{\i},
{\em Relativistic Quantum
Theory}\/ (Pergamon Press, Oxford, 1971).
\bibitem{Khrip}
I.B. Khriplovich, {\em Parity Nonconservation in
Atomic Phenomena}\/ (Gordon and Breach, New York, 1991).
\bibitem{Flambaum1976}
V.V. Flambaum, Yad. Fiz. {\bf 24}, 383
(1976) [Sov. J. Nucl. Phys. {\bf 24}, 199 (1976)].
\bibitem{KozLab95}
M.G. Kozlov and L.N. Labzowsky, J. Phys. B {\bf 28}, 1933 (1995).
\bibitem{KozEz94}
M.G. Kozlov and V.F. Ezhov, Phys. Rev. A {\bf 49}, 4502 (1994).
\bibitem{HLS76}
E.A. Hinds, C.E. Loving and P.G.H. Sandars, Phys.
Lett. B {\bf 62}, 97 (1976).
\bibitem{FlKhJETP1985}
V.V. Flambaum and I.B. Khriplovich, Zh. \'{E}ksp. Teor. Fiz.
{\bf 89}, 1505 (1985) [Sov. Phys. JETP {\bf 62}, 872 (1985)].
\bibitem{VF}
V.V. Flambaum, Doctor of Science Thesis, Novisibirsk, 
1987 (unpublished).
\bibitem{CSH91}
D. Cho, K. Sangster and E.A. Hinds,
Phys. Rev. A {\bf 44}, 2783 (1991).
\bibitem{Commins1994}
E.D. Commins, S.B. Ross, D. DeMille and B.C. Regan,
Phys. Rev. A {\bf 50}, 2960 (1994).
\bibitem{Smith1990}
K.F. Smith {\em et al.}\/, Phys. Lett. B {\bf 234}, 191 (1990).
\bibitem{Hg1993}
J.P. Jacobs, W.M. Klipstein, S.K. Lamoreaux, B.R. Heckel
and E.N. Fortson, Phys. Rev. Lett. {\bf 71}, 3782 (1993).
\end{thebibliography}
\end{document}